# Energy personas in Danish households

Nadine Sandjo Tchatchoua; Roskilde University, nadinet@ruc.dk

& Line Valdorff Madsen, Aalborg University, lvm@build.aau.dk

& Anders Rhiger Hansen, Aalborg University, arhansen@build.aau.dk

# Energy personas in Danish households

**Abstract.** Technologies to monitor the provision of renewable energy are part of emerging technologies to help address the discrepancy between renewable energy production and its related usage in households. This paper presents various ways householders use a technological artifact for the real-time monitoring of renewable energy provision. Such a monitoring thus affords householders with an opportunity to adjust their energy consumption according to renewable energy provision. In Denmark, EWII (previously Barry) is a Danish energy supplier which provides householders with an opportunity to monitor energy sources in 'real time' through a technological solution of the same name. This paper uses EWII's provision as a case for exploring how householders organize themselves to use a technological artefact that supports the monitoring of energy– and its related usage. This study aims to inform technology design through the derivation of four (4) personas. The derived personas highlight the differences in households' energy monitoring practices and engagement. These personas are characterised as 1) dedicated, 2) organised, 3) sporadic, and 4) convenient. Understanding these differences in energy monitoring practice using the technological artefact form a solid element in the design of future energy technologies that interfere with households' everyday practices and energy consumption. This is paramount for future energy related technology design, and for the clarification of usage assumptions that are embedded in the rollout of energy related technology as Denmark moves through its green transition.

**Keywords:** Energy, Energy monitoring technology, energy monitoring personas, household emerging technologies, technology design,

## 1. Introduction and Background

Electricity accounts for around one fifth of the total energy consumption worldwide, and therefore, the demand for electricity plays an important role in reducing CO2 emissions to meet political targets (OECD,2022). In Denmark, households consume one third of the energy produced, and although renewable energy production seem to increase substantially, households' total electricity consumption has increased with 11.5% from 1990 to 2020 (Danish Energy Agency 2021: Energy Statistics 2020, Danish Energy Agency, Denmark). To help with the decarbonisation of energy systems, technological solutions like the app-service 'Barry', which is the case of this study, provide a better information on real-time renewable production and tariffs. This also includes how and when householders use electricity in everyday practices and how energy suppliers interact with their customers. Similarly, online tools, smart metering and intermittent energy production underpin a new reality with new opportunities for energy suppliers and consumers (see for instance, Darby (2010), Littlechild (2021), and Große-Kreul (2022)).
Several technologies have been developed with the aim to provide more and better information to residential consumers. These are technological solutions that aim to motivate householders towards reducing their energy consumption altogether; or technological solutions that help time-shift energy consumption to periods with higher shares of renewable energy provision (Darby 2010).

Demand response initiatives to help balance energy production and consumption are central in the transition to a greener energy system and related tariffs is one way to engage households in flexible energy consumption practices (Torriti 2015). In several countries, including Denmark, such dynamic energy sources, where tariffs vary across the day depending on renewable energy availability on the grid, like Time-of-Use (ToU) pricing, have recently entered the electricity retail market with following discussions on how to design and develop these products (Hansen et al. 2022). However, successful demand response initiatives require a better understanding of energy demand and energy monitoring, as well as an understanding of the variations across household types (Torriti 2015; Hansen et al. 2022) and flexible building systems and design (Li et al. 2017).

In attempting to remedy to the lack of understanding of organisational structures linked to energy consumption in the household, researchers have acquired a sense of urgency into paying special attention to households as a unit of analysis for energy consumption (Adil and Ko 2016). Analysis on energy consumption has been in the fore of research in various disciplines (Motlagh et al. 2015), but with rapid transformations of energy markets and systems, more evidence is needed on how households respond in different ways to new situations.

Some studies have analysed technological interventions to suggest ways to balance energy supply and demand at the household level (Karnouskos et al. 2012), while others have attempted to segment households based on their consumption characteristics (Kamrat 2001). Within the social sciences, research on household energy consumption is in a large part built on social practice theory (SPT) (Gram-Hanssen 2010; Strengers et al. 2011; Madsen and Gram-Hanssen 2017). The SPT perspective focus on energy demanding practices that are shared across space and time and that energy is consumed in the course of accomplishing social practices (Shove et al. 2012, 2015; Shove and Walker 2014). Rather than focusing on technological solutions as such to change household energy consumption, often unrecognised smaller changes in routine are given more attention (Neustaedter et al. 2013), together with the materially and socially embeddedness of energy practices (Sahakian and Wilhite 2014; Jacobsen and Hansen 2021). In this sense, information cannot just be provided to households to increase knowledge and awareness, but surprising situations might underpin reflection, for example in times of disruption and crisis (Trentmann 2009).

In research on energy consumption in Danish households, most attention has been paid to heating, as this forms the major part of energy consumed (Madsen 2018). Also, increasingly there has been attention to demand shift opportunities of heating in homes that are heated through a district heating system (Larsen and Gram-Hanssen 2020).

Household electricity consumption has been somewhat overlooked within research on demand shift in Danish households, although there are examples of studies concerned with flexible consumption of electricity. For instance, Tjørring et al. (2018) investigated the flexibility of electricity consumption in private Danish households. An increased flexibility in electricity use is dependent on households and users that are actively engaged in planning their use of energy according to signals from the smart grid. For this purpose, technologies for energy management are needed and the active engagement of users in these technologies are dependent on a technical design that takes the user into account, for example based on everyday practices of households (Skjølsvold and Lindkvist 2015).

This paper studies energy monitoring practices of households based on the usage of a Danish technological intervention that allows Danish householders to monitor energy. The paper is based on qualitative interviews with Danish electricity customers, who have chosen a company they deem to have a 'greener' profile.

The analysis focuses on householders' accumulated experience to show how technology can reconfigure energy monitoring practices and to posit that technology has a key place in household energy monitoring support. Inspired by Schatzki (2009), we do so by deriving energy household monitoring personas from the 'doings and sayings' of householders to answer the research question 'In what ways could technology support energy consumption?'. Primary data from 14 semi-structured research interviews was collected by the first author. The empirical data we analyse has emerged from the use of an energy monitoring technology, designed by Barry (now EWII), in the said households over the course of over a year. We use rigorous data analysis and coding to extract four types of household energy monitoring personas in Denmark. These personas are intended to support energy policy and technology development for better future renewable energy consumption. Our main contribution is to highlight the importance of previous experience in energy monitoring practice using technology. We observe that there is not much existing literature based on social practice theory which attempts to classify householders for greener energy consumption purposes. This oversight is owned to the fact that social practice theory has historically focused on the collective of practices rather than examining the variations within the said practices. We explicitly address this gap by classifying household personas based on an analytical framework inspired from SPT.

Within the field of Computer supported Cooperative Work (CSCW), researchers (Schmidt and Bannon 1992, p. 13) have attempted to create a thread from artefact to infrastructure to lift off the 'interdependence in work' (see for instance Montero et al). Others have showcased a link between IOT data and energy data (Fischer et al. 2017). Finally, Wulf et al (2011) have experimented with creating design case studies. This paper takes inspiration from studies such as all the above to present household energy consumption persona data to contribute to the field of CSCW by supporting future green energy technology monitoring design. This paper aims to create a discourse around user household model candidates for specific green energy monitoring technology.

In what follows, we will attempt to contextualise and present personas based on previous literature and introduce this concept through the SPT framework. We will next give an account of the methodology used to derive our household energy monitoring personas. The next section will introduce our results based on the four derived household energy monitoring personas, followed by a discussion section in which we will argue that both technology and previous experience – of using technology - have a valuable place in energy monitoring support. To conclude, we summarise this paper by supporting that a focus on household energy practices can have a strong place in future technological and policy development. In doing so, we see these personas to have a positive impact in energy policy development to support the global green transition.

# 2 Analytical framework

## Social practices as an analytical lens for energy consumption

Social practice theory (SPT) has been widely used as an analytical approach for scrutinising and capturing dynamics of consumption, for example consumption of energy, and furthermore as a possible approach to interventions for design (Pettersen 2015). Thus, within the social sciences, SPT has gained a considerable interest in explaining energy use demanded by everyday practices. In this approach, it is understood that energy is not used for its own sake; rather, energy is used to accomplish social practices such as everyday activities in the home (Shove and Walker 2014). In this way, the demand for energy is understood as imbedded in the performance of collective and shared practices instead of a result of individual behaviour and decision-making. The demand for energy for heating and electricity is largely an invisible part of everyday life (Shove and Warde 2002). For instance, Gram-Hanssen (2010) demonstrates how social practice theory can be useful for understanding continuity and change in households' energy consumption, and a study involving interviews with 14 residential flats in Bergen in Norway show that residential practices influence energy consumption patterns (Wågø and Berker 2014). In the Norwegian study, householders were promised not to be required to change their energy consumption practices – given the way in which the houses were built; this soon proved to be unrealistic (ibid). Through an analysis of ten in-depth interviews with families participating in a project aimed at reducing standby consumption, Gram-Hanssen (2011) showed how technological configurations, everyday life routines, knowledge, and motivation constitute the practice and also structure the possibilities for change – towards better energy consumption.

Similarly, in an article aimed at highlighting practice from how people repair their broken objects, Wakkary et al. (2013) present practices involved in such repairs using Shove's (2012) materiality, competences and meanings framework including the interaction between such elements. The authors argue for a practice-oriented approach to help designers do away with classes of individual behaviour and studies of artefacts towards viewing sustainable behaviours as part of 'multidimensional' and interrelated practices. They posit that designers design towards resources and tools in ways that express the challenges of 'intelligibility' of their design interventions in practices. Moreover, they support that understanding the dynamics of practices and their unique configuration is a point of departure for the redefinition of roles in sustainable interaction design. It is to this effect that Ellegård and Palm (2011) urge designers to consider the analysis and understanding of energy consumption in relation to households' activity patterns as vital for developing policy means that contribute to an energy efficient life and what people would deem as a "good" everyday life.

This enables a shift from a sole focus on individual consumers to the broader reconfiguration of energy practices, such as energy monitoring, which is formed by technologies and other material arrangements, social conventions and expectations, and embodied experience and competences.

## Practice theoretical focus on households' personas for energy monitoring

'Persona' means mask and it is a concept that is widely used in the design industry (Nielsen 2019). Personas were first developed as a tool to support software developers in enabling them to better understand their target users (Cooper 1999). It is evident from the scarce literature on this topic that there is a lack of analysis of data that describes a specific range of existing households within the context of energy monitoring. That is, from a practice theoretical point of departure in Denmark. However, there has been examples of categorising householders, for example Larsen & Gram-Hanssen (2020) group users of smart home technology for heat management in households by ideal typical categories of user competences and engagement. Also, Madsen et al. (forthcoming) categorised ways of adapting new smart home technology in households and in this way grouped different types of SHT users, based on embodied competences to show the variety of SHT practices that should be met when SHTs are rolled out across a broad range of households to pursue a green transition of energy consumption and production. What is still lacking within the SPT literature is a model that can help guide designers in developing solutions that meet Danish households' needs. Given the fact that it is not practical to provide customised solutions for every Danish household, identification of needs for groups of similar households does provide a valuable insight and this paper offers a possible way forward and a solid baseline using personas.

The households' personas presented in this paper attempt to provide a robust support in the context of energy monitoring with technology in Denmark.

## 3   Research method

This paper is based on data collected as part of a wider project to understand how technology may support energy monitoring as a practice in Danish households. The interviews were conducted among 14 Danish households based in the greater Copenhagen area. The study informants were targeted from a Facebook group of households who have already made a leap towards renewable energy monitoring, and therefore could be referred to as early adopters (Rogers 2003).

### 3.1 Barry (now EWII)

At the time of the data collection, Barry was an application from an energy provider from the same name. On their website (https://barry.energy/dk/about, accessed 24 Jan 2022, own translation), Barry described themselves as a Finish company which generates and sells electricity and heat. Barry expanded to Denmark in 2018 and used the following catch phrase on their website *'A promise of "100% green power" is 100% hot air. But you probably already know that. On the other hand, we believe that data can show us the truth and help us on our way to a better everyday life and a greener future. We do this by showing you exactly how green your electricity is and how much $CO_2$ it emits, hour by hour - kWh for kWh.'*

Based on their core business model statement above, Barry as an energy provider focused on their ability to provide customers with the ability to see when the energy, they are consuming at the time is green. By 'green energy' they mean energy from renewable sources. The figure below, from Tchatchoua et al (2023) highlights the main affordances from Barry as an application. Barry became part of EWII in 2022. Please note that for the remainder of this paper, when we mention 'green energy', we will be referring to energy from renewable sources such as wind to be consistent with Barry's value proposition.

As can be seen from the screenshots below, provides real time energy source display as well as price and CO2 data. Barry affords its users the ability to wait for periods when the energy is from renewable sources to use energy.

*Figure 1 - Some of the features afforded by Barry as an energy monitoring technology.*

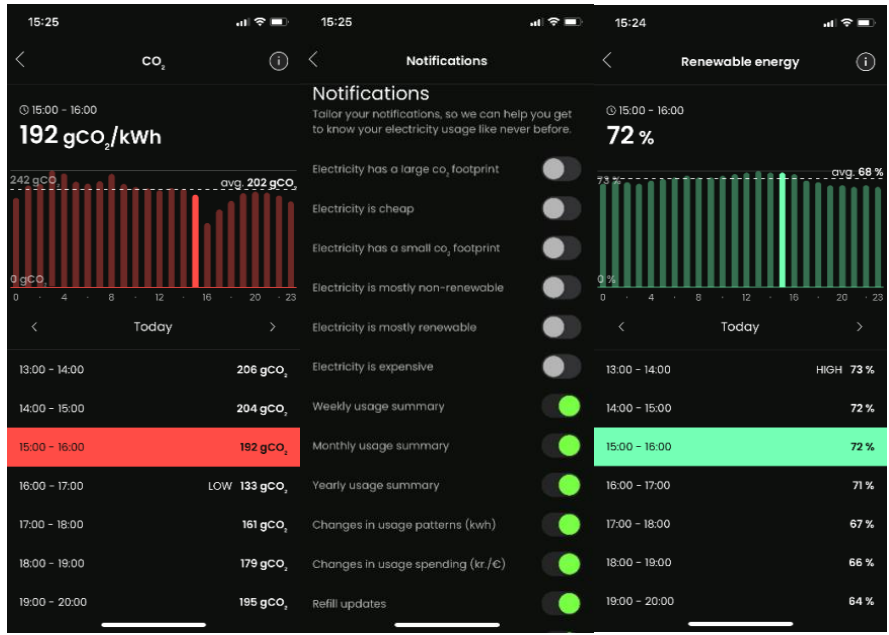

## 3.2 Data collection

In this project, we wanted to focus on specific energy monitoring practice scenarios and their special features – or contextual setting – for each of our 14 households. In doing so, we also aimed to dig out implicit aspects linked to the technology that affords the ability to monitor the said renewable energy. Observations in households had been planned initially but turned out not to be feasible in the end due to the ongoing pandemic. Online interviews turned out to be the sole data source for this research project. We were inspired by Schatzki in perceiving 'language as an important clue as to which activities and practices exist' (Schatzki 2009, p.9). Therefore, we found interview data to be valid and fitting when aggregating practice experiences.

The guiding research question was 'In what way can energy monitoring technological tools support energy monitoring practices of Danish households?', and because of its qualitative approach, we chose in-depth interviews as that method to conduct householders' experiences, knowledge, ideas and impressions (Alvesson 2003; Bryman 2016). In doing so, we aimed to be able to better influence future energy monitoring technology design.

We recruited household informants through snowball sampling and advertisements on Facebook, Twitter, and LinkedIn. These were householders using a new (at that time) energy monitoring tool from an energy provider in Denmark (called Barry). After suitable vetting by the group administrator, the corresponding author got in touch with Barry's group of users on Facebook.

The corresponding author conducted semi-structured interviews with participants representing individual households. Interviews typically lasted 30 to 60 minutes. Given the pandemic distancing measures at the time, most of the interviews had to be conducted online. However, two interviews were conducted physically, during a small window which allowed for direct interaction. The Interviews covered five principal themes: first, how and why the energy monitoring artefact was introduced into the household and its main usage norms within the household (e.g., who uses it, when and how). Next, the focus was shifted to the technological affordances that the energy monitoring artefact provides to the households, and the influence that it has (or has had) on householders' day to day activities. The informants also described the classic energy consumption routines in their households as well as a clarification of their homes' occupants, their energy consuming devices - including electronics such as mobile phones.

*Figure 1 Snippet Data Classification from NViVO*

| Household c | Found the gr | Being green | The green IT | Convenience | Introduced tl | I use the gre | Length of us | I work with t | How much d | I dont mind | Age range | My main dw | We live outs | Electric car |
|---|---|---|---|---|---|---|---|---|---|---|---|---|---|---|
| Cases\\Chris Married with | Yes | watching ou | Yes | Yes | Yes | Sometimes | Over a year | Yes | Not much th | Yes | 40-50 | Yes | No | Yes |
| Cases\\Jesp Married with | Yes | Living a bett | Yes | Unassigned | Yes | No | Less than a y | No | Regularly | Yes | 40-50 | Yes | Yes | Yes |
| Cases\\Osca Single Occup | Yes | Flying less a | Yes | No | Yes | Sometimes | Over a year | No | Not much th | Yes | Under 40 | No | No | No |
| Cases\\Chris Married with | Yes | Using energy | Yes | Unassigned | Yes | Yes | Over a year | No | Not much th | Yes | 40-50 | Yes | Yes | Yes |
| Cases\\Thon Married with | Yes | Looking afte | Yes | Unassigned | Yes | No | Over a year | Yes | Unassigned | Yes | 40-50 | Yes | Yes | Yes |
| Cases\\Kristi Married with | Yes | Living a bett | Yes | Unassigned | Yes | Sometimes | Over a year | Yes | Not much th | Yes | 40-50 | Yes | Yes | Yes |
| Cases\\Rasn Married with | Yes | An overwhel | Yes | Unassigned | Yes | Sometimes | Less than a y | No | Regularly | Yes | 40-50 | Yes | Yes | Yes |
| Cases\\Henr Married with | Yes | An overwhel | Yes | Unassigned | Yes | No | Over a year | Yes | Regularly | Yes | 50 + | Yes | Yes | Yes |
| Cases\\Nis Single with k | Yes | Looking afte | Yes | No | Yes | Yes | Over a year | Yes | Regularly | Yes | 40-50 | Yes | Yes | Yes |
| Cases\\Thon Married with | Yes | An overwhel | Yes | Unassigned | Yes | No | Less than a y | No | Regularly | Yes | 40-50 | Yes | Yes | Yes |
| Cases\\Sune Married with | No | An overwhel | Yes | Yes | Yes | Sometimes | Over a year | Yes | Not much th | Yes | 50 + | Yes | Yes | Yes |
| Cases\\Ande Single Occup | Yes | Looking afte | Yes | Yes | Yes | Sometimes | Over a year | Yes | Regularly | Yes | 40-50 | Yes | Yes | Yes |
| Cases\\Jesp Married with | No | Not Applicab | Yes | Unassigned | Yes | Unassigned | Over a year | Yes | Not much th | Yes | 50 + | Yes | Yes | No |

## 3.3 Data Analysis

The Interviews were conducted in English, recorded, and later transcribed. The corresponding author also took handwritten notes during the interview. We transcribed the recorded interviews and the handwritten notes helped to confirm the transcribed notes. The empirical data was coded using NVivo, through open, axial and selective coding (Corbin and Strauss 2008). In doing so, we aimed to bring to the fore resemblances and differences in energy monitoring practices - including in the usage of the energy monitoring technological artefact. Both induction and deduction were used when connecting and generating ideas from the empirical data. Induction was first applied in order to outline themes and patterns in the empirical data such as energy monitoring routines (Glaser and Strauss 2017). A deductive approach then followed when applying a practice theoretical lens to our empirical data, with a focus on the householders' learning from the experience of using the energy monitoring technological application.

When working inductively, the existence of certain qualitative features in a sample implies the presence of other features. To this effect, through induction, our study assumed that the participating households were a small instance of a known order. That is a bigger order (cohort) of householders in Denmark who have already adopted technology to support them in their energy monitoring, which could be referred to a group of early adopters (Rogers 2003).

Deduction was adopted when following the guiding research question more closely. Some of the features of the research context were already familiar to the first author and formed the basis for the subsumption process applied in the deduction process. The goal of the analytical approach was to understand each household's energy monitoring practices as well as the basic everyday energy monitoring routines - with technology.

### 3.4 Personas' derivation

The energy monitoring household personas in this paper are intended to support the requirements for the development of a wide spectrum of energy monitoring technologies and policies. This study's personas are developed with a focus on energy monitoring practices.
Since interview questions were open ended, certain attributes such as age and electric vehicle (EV) ownership were not part of the interview guide – and therefore not asked. These attributes were rather naturally volunteered by the informants during the interviews and included in the personas' specification.
Next, our results take us through the various stages of our analysis and present the four energy monitoring household personas we constructed from our empirical material. These personas have emerged from a point of departure where technology is used to support energy monitoring in our informants' households.

## 4   Results

The four energy monitoring personas presented in this study are intended to contribute to processes of design and implementation of future (green) energy monitoring technologies in line with the CSCW tradition. As we witness emerging technologies increasingly enter households (Hargreaves and Wilson 2017; Strengers and Nicholls 2017; Strengers et al. 2022), it is anticipated that such personas provide insight into the design of everyday energy monitoring technologies.

### The blue persona: *the dedicated household*

This model of households consists of a single occupant. This household would have been using the energy monitoring technology since the beginning of the start-up. To this effect, they feel part of the 'start-up family'. They find energy monitoring enjoyable as a practice. This household likes to 'play' with gadgets and their everyday energy consumption practices are quite set because of fewer disruptions in the home (no one else living with them).
In this household, the meal cooking practice is non-flexible. This means comfort comes before price as far as meal cooking is concerned (Tchatchoua et al. 2023). The household uses the energy monitoring technology often enough to the extent of finding the notifications redundant and disruptive. In addition to this, the household is keen to monitor energy to the extent that it wishes energy consumption data was displayed in 'real time'- the lag in data display is agonising here.
On the other hand, this household does not consider itself as making green transition related choices per se. The householder does believe however that monitoring energy through technology takes him/her a step closer in that direction. That is, through deciding to use energy during peak renewable energy availability periods. This household typically washes clothes once a week and puts the dishwasher on when it is full. Again, due to the limited disruptions in the household, it is easier for this household to wait for renewable energy availability before doing the washing for instance. The dedicated energy monitoring household's wish list for the app is a possibility to compare its renewable energy usage with that in a similar household.

### The green persona: *the organised household*

This household contains two adults and kid(s) still living at home. They have been using the energy monitoring technology since over a year ago although the male adult remains the sole user of the energy monitoring technology. In this household, the energy monitoring technology is not used every day, but only when the electric vehicle (EV) needs charging, i.e., three to four times a week. There exists a clear division of roles between genders in this household. This means that the male adult oversees the 'digital housekeeping' while the female adult takes care of the more traditional housekeeping routines (see Tchatchoua et al. (2023)).

Moreover, this household's everyday energy monitoring practices are a bit irregular since the children are 'too young to care about electricity consumption'. To this effect, this household doesn't look at the energy monitoring technology every day but has learnt to schedule certain energy consuming activities – such as the dish washing- to run during the night. This is, based on renewable energy availability information previously acquired from the energy monitoring technology. As in the dedicated household, meal cooking times are also set and independent of renewable energy availability here – due to personal preferences and bedtime routines. One of the households in this category we spoke to conceded: 'evening cooking times are quite set because of the child's bedtime'.

Moreover, in this household, the energy monitoring technology notifications do have a positive influence on energy monitoring behaviour - although they can't always act on them. The householders like the renewable energy data transparency offered by the energy monitoring technology. A householder in this persona observed: '…following sustainable routines is about being cautious when using energy'. They would rather not have energy monitoring activities 'take over'.

Finally, for this household persona, clothes are washed every day and the dishwasher is on twice a day – mostly during the night as aforementioned. That is, based on the learning about renewable energy availability periods from the energy monitoring technology. The organised energy monitoring household's wish list for the app is a possibility to integrate several energy sources – both from digital and analogue sources - in one technological artefact.

### The red persona: *the sporadic household*

This household consists of a single occupant, who is also a parent sharing custody of (a) child(ren). This household has been using the energy monitoring technology since the acquisition of a new EV. The adult is also the sole energy monitoring technology user in this household. The children's sporadic presence means they don't get involved in energy consumption matters much. This also means that the household everyday energy monitoring practices are sporadic. This household however tries to use the energy monitoring technology every day and finds the notifications to be a useful reminder on days when they may forget to use it. This household is quite flexible in shifting its renewable energy consumption routines but do however deem cooking a non-negotiable energy consumption practice. One of the participants in this persona type of household strongly exemplified this with the following statement: 'I am not going to be influenced by energy price for cooking'.

For this household persona, having to follow sustainable routines is about taking care of the planet and using less water for instance. They typically wash clothes twice a week and charge the electric vehicle once a week on average. The sporadic energy monitoring persona type would like the energy monitoring technology to advise on better ways to use renewable energy - from previous renewable energy consumption data. They explain that it would mean the technology working with the following assumption '[…] you usually use a lot of green energy in this timeframe when the price is quite high, perhaps you could shift some of that to other times for example'. The energy monitoring technology could similarly also recommend when to best charge the EV for instance.

### The yellow Persona: *the convenient household*

In this household, there are two partnered adults and kid(s) living at home permanently (as for the green persona). Here, the energy monitoring technology has also been used for quite some time. What is striking about this household persona is that both the adults use the energy monitoring technology with a similar level of engagement.

However, the general level of interest in the energy monitoring technology appears to have faded slightly for both adults. For example, a householder observed: '[…] in the beginning it was kind of fun it is not something we use actively now'. The reticent sentiment in such a statement highlights this household's everyday energy monitoring practices, which are now are sporadic, since they typically choose convenience over price, regardless of renewable energy availability. They have disabled the energy monitoring technology notifications, but they would paradoxically postpone charging the electric vehicle if the only energy available was not from renewable sources.

Households in this energy monitoring household type reported 'we no longer find it -energy monitoring using the technological artefact- fun'. Both adults in this household have used the energy monitoring technology -on equal terms – for long enough to memorise frequent periods of renewable energy availability. 'I would never run the washer from around 4pm in the afternoon until 8pm because we know for a fact that that is when most people use energy. And that is when it is most expensive and not from renewable sources'. To this type of households, backing sustainable routines is about trying to live sustainably by washing clothes fewer times during the week for instance. Interestingly, they typically use the dishwasher everyday as a practice related to cooking, that they enjoy doing and that is also a steady practice.

The yellow energy monitoring household would like the energy monitoring technology to better integrate with smart home systems instead of having disjointed energy related systems throughout the home.

## 5 Discussion

This study uses a Danish energy supplier - which also provides an energy monitoring technological application – to present four personas reflecting different ways of engaging in and practicing energy monitoring. Such a technological artefact provides new opportunities for householders to engage in energy monitoring practices. This analysis has illustrated how similar customers (and users) approach such a practice in different ways, while also emphasizing how renewable energy monitoring practices are related to the socio-technical context. Such an emphasis is to be considered in the design of adequate technology to enhance renewable energy monitoring, due to its sporadic nature, as Denmark moves into its green transition.

The personas derived in this paper are based on the methodological approach outlined by Nielsen (2019), which consists of writing short stories based on interviews with users of technology or software. Our results paint a concise picture of each of the household personas to present ways they can be supported in their renewable energy monitoring initiative through technology and policy design.

Table 2 summarises the four energy monitoring household personas empirically defined in this study. They present differences in terms of their engagement in monitoring renewable energy supply – and its related usage - as a practice. They also have divergent views on what it takes to be 'green' as a household. Finally, they offer diverse opportunities for future energy monitoring technology design.

In particular, the personas of this paper will give insights around: 1) energy monitoring practices in households, 2) the socio-technical challenges related to renewable energy monitoring at the household level, 3) energy monitoring technology usage in the household, and finally, 4) renewable energy monitoring as a practice, resulting in opportunities and -or- barriers for better energy monitoring technology and policy design.

What all these household personas have in common is usage of the energy monitoring technology acting as an element holding their renewable energy monitoring – and its related usage- practices together. That is, regardless of the level of interaction with the said technology (see Tchatchoua et al. 2023). It is important to acknowledge at this stage that these energy monitoring household personas are rather similar by means of socio-economic characteristics. They are all based in the greater Copenhagen region. They also share similar technological competences and motivation, for example by being early adopters (Rogers 2003) with a keen interest in technology. For them, the energy monitoring technology is a 'fashionable object of desire' (Pantzar 1997). A key element in these personas is householders' 'wish list' for future energy monitoring technology design. Finally, these personas are to be perceived as typical representations of different energy monitoring households.

*Table 1 Overview of household personas.*

| Household persona | Engagement in energy monitoring as a practice | What it means to follow sustainable routines as a household | Opportunity for future technology design |
|---|---|---|---|
| Blue | Dedicated and playful: Using energy monitoring technology is enjoyable | Energy monitoring is one step closer to following sustainable routines. | Comparison of renewable energy consumption with similar households. |
| Green | Organised: Engagement with the energy monitoring technology is irregular yet practical. | Following sustainable routines is time-consuming and needs to fit into other everyday routines. | Integration of several energy sources (digital and analogue) in the same technology. |
| Red | Sporadic: Engagement with the energy monitoring technology varies a lot. | Following sustainable routines is about taking care of the planet and using less energy. | Include advice on better ways of using energy from renewable sources, for example based on historical energy consumption data. |
| Yellow | Convenient: Comfort and what makes everyday practices function come first. | Following sustainable routines is about trying to live more | Better integration with smart home systems. |

| | | sustainable by for example washing less. | |

Overall, across the four personas, there were little interest in prices as an incentive for energy monitoring their everyday household practices[1]. The primary reason for monitoring renewable energy availability was for charging EVs and time-varying prices was not perceived as a primary motivator for time-shifting additional household energy consumption practice. Such practice has obvious implications for technology and policy design measures that typically focus on price reduction as a sole motivator.

Furthermore, the personas presented in this paper highlight that those householders, despite the same cultural and geographical settings, do vary in their socio-material settings when it comes to renewable energy monitoring – and its related usage. Thus, different practices are to be considered in energy monitoring technology design. To some extent, these resulting personas resonate with the scarce findings on personas from the practice theory literature as aforementioned in the introductory section of this paper.

On the other hand, it is clear from our results that all household personas appreciated the renewable energy data visualisation affordance (see fig 1) provided by the energy monitoring technology. In other words, all four personas appreciated – albeit initially for some - the feedback they received from the technologies. The dedicated energy monitoring household persona demonstrates this most strongly, reflecting that energy monitoring is an experience as well as a task, placing a high priority on the experience or skill. To this effect, Shove et al. (2012) highlight the value in developing competences in practices.

The dedicated energy monitoring household provides a real opportunity for policy makers and developers. The provision of subsidised renewable energy tariffs coupled with innovative technological solutions to support SDGs should offer the opportunity for playful energy monitoring households to shift their energy demands even more. This is owned to the fact that some of these households are motivated – and have capacity - to put up with the disruption and effort energy monitoring as a practice requires.

The organised energy monitoring household takes after the idea of 'the resource man' developed by Strengers (2014). Here, there is an existing aspiration to monitor energy and thus align renewable energy consumption practices. However, the organised energy monitoring household is characterised by having one main actor who interacts with the technology and is in charge of what (Tolmie et al., 2007) describe as digital housekeeping or Martin describes as energy housekeeping (Martin 2022). Meanwhile, the female adult in the household still takes charge of the more traditional type of housekeeping. One of the informants in this type of households conceded: 'I tell my wife what to do and she does it'. They were referring to the best times to put the washing machine on according to renewable energy availability slots by this statement. Similar examples of gendered divisions of household labour and digital/energy housekeeping have also been shown in a paper by Martin (2022) as well as within other studies (Johnson 2020; Aagaard 2022; Aggeli et al. 2022).

Opportunities for energy monitoring technologies' design would only be viable if the household dynamics can be changed to allow for both adult actors to take part in the energy monitoring practice.

---

[1] Please note that this data was gathered before the current energy crisis and this sentiment may have changed because of the said crisis.

Similarly, the sporadic energy monitoring households also want to improve their energy monitoring practice as a household, however, their everyday setting is not conducive of routinised practices. That is, in line with the current renewable energy monitoring – and related usage - routine that the technology advocates. To these types of households, energy monitoring technology is a support that needs reminders for its very usage. This household persona relies heavily on the experience from previous usage of the energy monitoring technology for continuous energy monitoring practice during busy periods.

Finally, for the convenient energy monitoring household persona, there exists a more gender-equal energy monitoring approach between the partnered adults. That is, in these types of households, adults in the household tend to possess a similar motivation and aptitude to use the energy monitoring technology to support their renewable energy monitoring and consumption practices. Both the adults in this type of household persona appear to work as a team when interacting and communicating their preferences in relation to energy monitoring to one another. In effect, this household persona finds the energy monitoring technological artefact to be a solid foundation in their renewable energy monitoring and consumption routines; this includes their decision making as a team. Close energy monitoring communication unfortunately also means that when one adult becomes reluctant to use the technology, they might influence the other one to follow suit. Intuitive technology design with a mixed cohort of stakeholders would prove suitable for the development of technology and policies for such households.

The gendered differences in households' use of smart energy technologies in the home has also been addressed by other authors, such as Aagaard and Madsen (2022), who show that in 'frontrunner' households it is often the male householder who is the engaged and competent user of the smart home technologies (SHT), while in non-frontrunner households the engagement and competence in using SHT can be more evenly shared. Thus, these gendered differences in and between households is central to consider in the design and implementation of smart home technologies and households' energy monitoring practices.

## 6   Scope for further research

This study presents a scope for further research and validation of four derived personas due to the fact the sample of informants was limited and quite homogenous. To exemplify this, 12 of the 14 households included in this sample were electric vehicle owners (Tesla), which was important for choosing Barry as an energy provider and the related technology to monitor energy consumption. Having stated that, homogenous sampling is not foreign in energy-related studies – see for example Wunderlich et al. (2013). The study's focus was to capture householders at the forefront of the energy monitoring chain as a practice, so we could learn from them for future energy monitoring technology design for a more diverse group of users. We posit that this study offers a rich foundation for further investigations that could use a wider sample with a mixed cohort of informants and supplementary sources of data.

# 7 Design implications

The presented household energy monitoring personas advocate for customisable energy monitoring technological artefacts to meet various household specific demands. In addition to this, every household persona reported in this study typically has acquired the energy monitoring technology following the acquisition of a new electric vehicle. However, yet only half of the derived household personas (the dedicated and organised personas) present a keen enthusiasm for using the energy monitoring technology to support their renewable energy consumption.

Moreover, some of the household personas -although technically apt-, felt patronised by the energy monitoring technology notifications. In response to this, the households felt reluctant to alter certain practices they deemed necessary and non-negotiable energy consumption practices, for instance cooking. As a result, the householders would typically turn off all notifications from the energy monitoring technology. This behaviour could be explained by findings from a growing body of research that reveals how some households find imposed energy-saving policies demotivating - see (Ellegård and Palm 2011). We see these models of households as part of a nascent and an emerging body of studies that need particular attention, as post pandemic flexible working methods influence the way householders consume energy. This study is important as the world slumbers in the current energy and climate crisis with the war in Ukraine.

**Figure 3 Energy monitoring technology design implications for the four households' personas.**

**Dedicated**
- Keen to monitor energy.
- Could be offered new interfaces and products periodically in order to curve product familiarity and enhance the energy monitoring motivation.
- Willing and able to shift electricity demands.
- Would be a good candidate for new and innovative energy monitoring technology testing for feedback.

**Organised**
- Happy to monitor energy through the use of technology as long as it does not alter their household dynamics.
- There is typically one adult in charge of the energy monitoring routines in this type of household.
- Needs a customised energy monitoring technology, for example with settings that meet their household needs.
- The energy monitoring technology design needs to content fewer notifications and much less features for instance.

**Sporadic**
- The dynamics and change of routines do not allow for a regular pattern in renewable energy monitoring -and its related consumption- as such although there is interest in being consistent with the said practice.
- There are eclectic design opportunities with different energy source types offerings and tariffs here- to allow for the peaks and drops in energy monitoring practice.

**Convenient**
- The adults in this household persona type monitor energy as a team. They do also influence each other's decision to pursue such a practice.
- This household persona encourages an inclusive energy monitoring technology design since both adults in the household work as a team and are keen to engage with energy monitoring.

These four personas thus provide us with an insight into the motivations behind energy monitoring at the household level. In so doing, they create opportunities for engagement with renewable energy monitoring (and consumption) technology design within Denmark – and the world by and large.

Literature does explain that there remain households that will never engage with technology long term (Jensen et al. 2018). This is perhaps owned to the fact that their private circumstances mean they transition into different household settings frequently, hence them being unable to find a form of stability necessary for regular energy monitoring. Households occupied by the elderly may struggle with the ability to meet the demands of technology usage. To this effect, energy monitoring technology design is to allow for such households and find ways to afford them to partake in renewable energy monitoring and consumption activities regardless of the sporadicity of such energy. Similarly, renewable energy monitoring and consumption design measures could include different renewable energy tariffs to cater for and incentivise such households. The said measures could allow for renewable energy availability to be given priority to elderly householders for instance, thus leading to a default reduction in need to interact with energy monitoring technology in such households.

Finally, it is worth emphasising here that the household personas specified in this paper are subject to change over time - depending on their current social set up. That is, one household in a specific persona may move onto another with time depending on a new context and social circumstances – after a divorce for example. In other words, a dedicated energy monitoring household may become organised for instance – or vice versa. This persona type evolution could also to linked to a shift in focus in renewable energy monitoring routine. Successful energy monitoring technology and policy designs are thus to allow for flexibility to cater for these potential changes in social circumstances.

# 8   Conclusion

In this paper, we have offered the grounding for future energy monitoring technology design through the derivation of four household personas. Using energy practices as a lens to understand energy demand, we posit in this paper that, the concept of persona development in CSCW related studies is scarce yet important when designing technology to support energy monitoring in the home. We conclude that combining energy consumption studies and personas offer a new and unique opportunity in CSCW to support technology design. We therefore anticipate these households' personas to form a solid foundation for future energy monitoring technology research and the related renewable energy consumption through practice.